\documentclass[12pt,a4paper]{llncs}
\usepackage{amsmath,amssymb,amsfonts,bm,dcolumn,booktabs,nicematrix}
\usepackage[hidelinks,colorlinks=true,urlcolor=blue]{hyperref}
\usepackage{geometry}
\newcommand{\sft}[1]{\mbox{\tiny{#1}}}

\newcommand{\ds}[1]{$\displaystyle{#1}$}
\newcommand{\mpcb}[2]{\begin{minipage}{#1}\begin{center}{\bf #2}\end{center}\end{minipage}}
\newcommand{\mpc}[2]{\begin{minipage}{#1}\begin{center}{#2}\end{center}\end{minipage}}

\begin{document}

\title{Quantum equilibration and measurements -- bounds on speeds, Lyapunov exponents, and transport coefficients obtained from the uncertainty relations and their comparison with experimental data}
\author{Saurish Chakrabarty\inst{1}\orcidID{0000-0001-8534-8644} \and Zohar Nussinov\inst{2,3}\orcidID{0000-0001-7947-7829}}
\institute{Department of Physics, Acharya Prafulla Chandra College, New Barrackpore, Kolkata 700131, India\\ \email{\href{mailto:saurish@apccollege.ac.in}{saurish@apccollege.ac.in}} \and
  Rudolf Peierls Centre for Theoretical Physics, University of Oxford, Oxford OX1 3PU, United Kingdom\\
  \email{\href{mailto:zohar.nussinov@physics.ox.ac.uk}{zohar.nussinov@physics.ox.ac.uk}} \and
  Department of Physics, Washington University in St. Louis, 1 Brookings Drive, St. Louis, MO 63130,
  USA\\
  \email{\href{mailto:zohar@wustl.edu}{zohar@wustl.edu}}  
}
\maketitle
\begin{abstract}
  We discuss our recent study of {\em local quantum mechanical uncertainty relations} in quantum many body systems. These lead to fundamental bounds for quantities such as the speed, acceleration, relaxation times, 
  spatial gradients and the Lyapunov exponents. We additionally obtain bounds on various transport coefficients like the viscosity, the diffusion constant, and the thermal conductivity. Some of these bounds are related to earlier conjectures, such as the bound on chaos by Maldacena, Shenker and Stanford while others are new. Our approach is a direct way of obtaining exact bounds in fairly general settings. We employ uncertainty relations for local quantities from which we strip off irrelevant terms as much as possible, thereby removing non-local terms. To gauge the utility of our bounds, we briefly compare their numerical values with typical values available from experimental data. In various cases, approximate simplified variants of the bounds that we obtain can become fairly tight, $i.e.$, comparable to experimental values. 
  These considerations lead to a minimal time for thermal equilibrium to be achieved. Building on a {\em conjectured relation between quantum measurements and equilibration}, our bounds, far more speculatively, suggest a minimal time scale for measurements to stabilize to equilibrium values.
\end{abstract}

\section{Introduction}
In this work, we summarize our recent findings discussed in Refs.  \cite{planckianAOP,macroscopic} and briefly compare rigorous
bounds on physical quantities that we obtained using our approach with experimental data. 
A large number of conjectured bounds on physical quantities have been advanced.
These include
an upper bound on the Lyapunov exponent \cite{mssChaos},
a lower bound on various lifetimes and relaxation rates \cite{planckianBruin,Ramshaw,melting-speed,nnbk-viscosity},
a lower bound on the viscosity \cite{Anup,nnbk-viscosity,zoharViscosityPRR,jan-planck},
a lower bound on the ratio of shear viscosity and entropy density \cite{KSS2},
and many others. It is notable that early works by Eyring \cite{Eyring1,Eyring2} and other pioneers on chemical reaction rates and intuitive proposed extensions implicitly suggest similar inequalities (although these have not been proposed as fundamental bounds). 
Our primary goal is to rigorously derive such bounds in broad settings using local variants of the quantum mechanical uncertainty relations.

\section{Bounds from local uncertainty relations in many body systems}
We consider a macroscopic system $\Lambda$ of $N_{\Lambda}$ particles, with a density matrix $\rho_\Lambda$, whose dynamics is
governed by the time independent Hamiltonian $H_\Lambda$.
The rate of change of an arbitrary local operator $Q_i^H$  in the Heisenberg picture is 
\ds{\frac{dQ_i^H}{dt}=\frac{i}{\hbar}\left[H_\Lambda,Q_i^H\right]}.
The subscript $i$ can be thought of as a particle index.
We note that we can replace $H_\Lambda$ in the above expression by the local Heisenberg picture Hamiltonian $\tilde{H}_i^H$
which represents only the portion of $H_\Lambda$ containing terms that do not commute with our
chosen local operator $Q_i^H$. With this, 
\ds{\frac{dQ_i^H}{dt}=\frac{i}{\hbar}\left[\tilde{H}_i^H,Q_i^H\right]}.
Next, we use the textbook type quantum uncertainty relation which is trivially provable to be valid $\Big($via, $e.g.$, the use of Cauchy-Schwarz inequalities for Hilbert-Schmidt (trace) type inner products satisfying the inner product positive semi-definite property $\big({\sf Tr} (\rho_{\Lambda} A^{\dagger} A) \ge 0\big)$ associated with the density matrices $\rho_{\Lambda}$ providing expectation values$\Big)$ for general mixed states, 
\ds{\sigma_A~\sigma_B\ge\frac{1}{2}\left|\big\langle[A,B]\big\rangle\right|}.
Here, $A$ and $B$ are any two operators and
\ds{\sigma_A^2=\left\langle(A-\langle A\rangle)^2\right\rangle},
\ds{\langle A\rangle\equiv\mbox{Tr}\left(\rho_\Lambda A\right)}.
Using this, 
  \ds{\left|\left\langle\frac{dQ_i^H}{dt}\right\rangle\right|
  \le\frac{2}{\hbar}\sigma_{\tilde{H}_i^H}\sigma_{Q_i^H}}.
Now we focus on the value of $\sigma_{\tilde{H}_i^H}^2$ when averaged over the entire system and consider the particular case of $\rho_{\Lambda}$ defining a macroscopic thermal system at a temperature $T$ for which the variances may be evaluated. For a translationally invariant 
system in thermal equilibrium, the variance $\left(\sigma_{\tilde{H}_i^H}\right)^2 \equiv k_BT^2C_{v,i}$ (defining an effective local heat capacity  $C_{v,i}$) assumes the same value of each $i$. (The energy variance of the full many body Hamiltonian $H_{\Lambda}$ is given by $k_{B} T^2 C^{(\Lambda)}_v$ with $C^{(\Lambda)}_v$ the heat capacity of the global system $\Lambda$.) Putting everything together, 
\begin{eqnarray}
  \overline{\left(\frac{\left\langle\frac{dQ^H}{dt}\right\rangle^2}{\sigma_{Q^H}^2}\right)}
  \le\frac{4k_BT^2C_{v,i}}{\hbar^2},
  \label{rateSqBound}
\end{eqnarray}
\vspace{-7mm}~\\
where $\overline{X}\equiv\frac{1}{N_\Lambda}\sum\limits_{i=1}^{N_\Lambda}X_i$.
Even though the right hand side of Eq. \ref{rateSqBound} is independent of the spatial index $i$, we have kept it to underscore that $C_{v,i}$ is an effective {\it local} heat capacity.

\subsection{Upper bound on the relaxation rate (Lower bound on the relaxation time)}
The left hand side of Eq. \ref{rateSqBound} is, dimensionally, the square of the
relaxation rate associated with the operator $Q_i^H$. This leads to a bound on the
relaxation rate,
\begin{eqnarray}
  \tau_Q^{-1}
  \le\frac{2T\sqrt{k_BC_{v,i}}}{\hbar}.
  \label{rateBound}
\end{eqnarray}
At high temperatures, when the equipartition theorem applies, $i.e.$, 
$C_{v,i}={\cal O}(k_B)$, this inequality becomes,
$\tau_Q^{-1} \le{\cal O}\left(2k_BT/\hbar\right)$,
  implying that,
\ds{\tau_Q
  \ge{\cal O}\left(\hbar/2k_BT\right)}.  

\subsection{Upper bound on particle speeds and lower bounds on particle displacements}
Choosing the operator $Q_i^H$ in the above analysis to be the $\alpha^{\sft{th}}$ Euclidean component
of the displacement of a particle in the system, we get,
\ds{\overline{\left(\left\langle\frac{dr_\alpha^H}{dt}\right\rangle^2\Big/\sigma_{r_\alpha^H}^2\right)}
  \le\frac{4k_BT^2C_{v,i}}{\hbar^2}}.
Here, $\tilde{H}_i^H=\frac{\left(p_{i\alpha}^H\right)^2}{2m}$, implying that if equipartition holds
(at high temperatures), $C_{v,i}=k_B/2$. If in addition, we assume that the fluctuation of the particle
positions is slowly varying, $i.e.$, all the particles have similar values of $\sigma_{r_{i\alpha}^H}$,
then,
\begin{eqnarray}
  \sqrt{\overline{\left\langle\frac{dr_\alpha^H}{dt}\right\rangle^2}}
  \le\frac{\sqrt{2}k_BT\sigma_{r_\alpha^H}}{\hbar}.
\end{eqnarray}
A related bound for the expectation value of the square of the velocity components can also be
obtained using a similar analysis. \cite{planckianAOP}
Thus, at high temperatures,
\ds{\overline{\left\langle\left(\frac{dr_\alpha^H}{dt}\right)^2\right\rangle}
  \le\frac{2\left(k_BT\right)^2\sigma_{r_\alpha^H}^2}{\hbar^2}}.
The advantage of this relation is that in the classical limit, the left hand side takes the
value $\frac{k_BT}{m}$, implying that the fluctuation of the each component of a particle's
position is bounded from below.
\begin{eqnarray}
  \sigma_{r_\alpha^H}^2\ge \frac{\hbar^2}{2mk_BT}=\frac{\lambda_T^2}{4\pi},
\end{eqnarray}
\vspace{-10mm}~\\
$\lambda_T$ being the thermal de Broglie wavelength.

\begin{table}
  \begin{center}
    \vspace{-5mm}
    \begin{NiceTabular}{|c|c|c|c|c|}
      \toprule
      {\bf Quantity} & \mpcb{4cm}{Simplified bound\\(order of magnitude)} &
      \mpcb{25mm}{Approximate value of bound} &
      \mpcb{29mm}{Typical value(s) of quantity} &
      \mpcb{38mm}{Related conjectured bounds} \\
      \midrule
      \mpc{15mm}{Relaxation time} & \ds{\tau\gtrsim\frac{\hbar}{2k_BT}}
      & \mpc{25mm}{0.01 ps\\ ($T=300$ K)}
      & --- & Planckian time  \cite{zaanen2004}\\
      \midrule
      Speed & \ds{|v|\lesssim\frac{k_BT\sigma_{r_\alpha^H}\sqrt{2}}{\hbar}} &
      \mpc{25mm}{7 km/s\\ ($T=660^\circ$C,\\$\sigma_{r_\alpha^H}=0.4~\AA$)}
      & \mpc{38mm}{3 km/s (sound speed in aluminum at $660^\circ$C, just below
        its melting point)} &
      Melting speed  \cite{melting-speed}\\
      \midrule
      \mpc{15mm}{Diffusion constant} & \ds{D\gtrsim\frac{\hbar}{2\pi m}} &
      \mpc{3cm}{$5\times10^{-10}$ m$^2$/s\\(for water)}
      &\mpc{3cm}{$1.1\times10^{-9}$ m$^2$/s\\(for water at STP)}
      & --- \\
      \midrule
      \mpc{15mm}{Shear viscosity}
      & \mpc{3cm}{\ds{\eta\gtrsim nh}\\\ds{\eta\lesssim\frac{mk_BT}{3R\hbar}}}
      & \mpc{2.5cm}{$\eta\gtrsim2\times10^{-5}$ Pa s\\\vspace{-2mm}~\\
      $\eta\lesssim3\times10^{-3}$ Pa s\\(for water)} &
      \mpc{2cm}{$6\times10^{-5}$ Pa s\\\vspace{-2mm}~\\
        $2\times10^{-3}$ Pa s\\(for water)} &
      \mpc{42mm}{Bound on viscosity to entropy density  \cite{KSS2},
      Minimum viscosity  \cite{Anup,jan-planck,zoharViscosityPRR}, Minimum kinematic viscosity  \cite{KT,ktRev}} 
      \\
      \midrule
      \mpc{15mm}{Bulk viscosity} & \ds{\zeta\gtrsim\frac{n\hbar}{\sqrt{d^3(z+1)}}} &
      \mpc{3cm}{$3\times10^{-7}$ Pa s \\(for water at $100^\circ$C)} &
      \mpc{3cm}{$5\times10^{-4}$ Pa s\\(for water at $100^\circ$C)}
      & --- \\
      \midrule
      \mpc{15mm}{Lyapunov exponent} & \ds{\lambda_L\lesssim\frac{2k_BT\sqrt{d}}{\hbar}}
      & \mpc{3cm}{$10^{14}$ Hz\\($T=300$ K, $d=3$)} & --- &
      \mpc{35mm}{Chaos bound,\\ \ds{\lambda_L\lesssim\frac{2\pi k_BT}{\hbar}}  \cite{mssChaos}}\\
      \midrule
      \mpc{15mm}{Spatial gradient} & \ds{\overline{\left\langle\left(\frac{\partial f}{\partial r_\alpha}\right)^2\right\rangle\Big/\left\langle f^2\right\rangle}\le\frac{8\pi}{\lambda_T^2}}
       & --- & --- & ---\\
      \bottomrule
    \end{NiceTabular}
  \end{center}
  \caption{A summary of our bounds for quantum thermal systems at a temperature $T$.
    Here, $m$ denotes the mass of a particle,
    $R$ its radius, and 
    $\sigma_{r_{i\alpha}^H}$ is the standard deviation 
    of a Cartesian component ($\alpha$) of its position,
    $d$ is the number of spatial dimensions, 
    $n$ the number of particles per unit volume,
    $z$ the effective coordination number, $f$ an arbitrary function of the spatial coordinates and other degrees of freedom. 
    In this table, we used the classical equipartition theorem at sufficiently high temperatures where classical equipartition applies. Further simplifying approximations were made to our exact bounds. In several of our inequalities, the Green-Kubo type integrals for the diffusion constant and other transport coefficients were evaluated up to the first zero of their respective particle velocity and other autocorrelation functions and rigorously bounded (thus not accounting for localization effects ($D=0$) that appear when oscillatory autocorrelation function contributions are dominant). More physically transparent approximate derivations lead to some of the above shown inequalities. Comprehensive details on these bounds and those on other quantities
    can be found in Ref.  \cite{planckianAOP}. }
  \label{boundsTab}
\end{table}

Other bounds that can be obtained using similar analysis are summarized in
Table \ref{boundsTab}. These have been simplified using semi-classical
and other arguments in order to obtain expressions devoid of
specific system details.

\section{Quantum measurements and equilibration}
In Refs. \cite{planckianAOP,macroscopic}, the Eigenstate Thermalization Hypothesis, associated entropy maximization, and other considerations were applied to the measurement problem. Here, the interactions $H_{\sf device-Q}$ between a measuring device and a local microscopic quantity $Q$ being measured were included in the full system Hamiltonian $H_{\Lambda}$. It was illustrated that a time average of $Q$ (over its equilibration time set by $\tau_{Q}$) is given by {\it eigenstate expectation values} when the interactions in $H_{\sf device-Q}$ are appreciable. That is, inasmuch as local measurements of $Q$ are concerned  \cite{planckianAOP,macroscopic},
\begin{eqnarray}
\label{eq:eqcollapse}
\rho_{\sf collapse} \mbox{``=''} \rho_{\sf equil.},
\end{eqnarray}
where $\rho_{\sf collapse}$ is the density matrix associated with this short time averaged measurement and $\rho_{\sf equil.}$ emphasizes that the latter short time average may be replaced by an average with the density matrix of the equilibrated system that includes the measurement device and the typical microscopic quantity $Q$ being measured. Here, ``='' highlights that this equality and density matrix are not associated with a bona fide ``collapse'' to an eigenstate of $Q$ but rather to a time average over an interval which can be exceedingly short for a small local observable $Q$ (see Table \ref{boundsTab}) for which equilibration may indeed typically be very rapid. Ref. \cite{same_conjecture} more recently raised a conjecture similar to the one of Eq. (\ref{eq:eqcollapse}) that we earlier proposed in Refs. \cite{planckianAOP,macroscopic}.

\section{Conclusions}
Our local quantum uncertainty based bounds on the relaxation times  in equilibrated quantum systems \cite{planckianAOP,macroscopic} are intimately related to conjectured Matsubara like Planckian time
scales \cite{zaanen2004} and do not hinge on the Lieb-Robinson \cite{Lieb_Robinson} and related bounds \cite{taddeiQuantumSpeedLimit} on the speed in which information may spread. These bounds may further relate to possible fundamental limits on measurement and equilibration times (a conjectured connection between measurement and equilibration was briefly reviewed). Our lower bound on the shear viscosity is closely connected to proposed bounds on the viscosity to entropy density ratio  \cite{KSS2},
and other viscosity bounds  \cite{jan-planck,zoharViscosityPRR,KT}.
Our upper bound on the shear viscosity in equilibrated systems, that follows from the
bound on the diffusion constant when the Stokes-Einstein
relation applies is, like others reviewed here (e.g., those on general spatial gradients of general functions), new \cite{planckianAOP}. When applied to various observables, our bound on the Lyapunov exponent is slightly tighter than the celebrated conjectured chaos bound of Ref. \cite{mssChaos}. Furthermore, our derivation uses a definition of the Lyapunov exponent similar to that in the 
the classical arena which does not rely on the use of regularized Out of Time Ordered
Correlators (OTOC). When contrasted with experimental data for commonplace systems such as water and aluminum, our simplified bounds are relatively tight (see Table \ref{boundsTab} and \cite{planckianAOP} (and further comparisons for the viscosity bound in \cite{Anup,zoharViscosityPRR})). A comprehensive study further contrasting some of our other bounds (both exact and their approximate simplified variants) with experimental data will be illuminating.


\begin{thebibliography}{10}
\providecommand{\url}[1]{\texttt{#1}}
\providecommand{\urlprefix}{URL }
\providecommand{\doi}[1]{https://doi.org/#1}

\bibitem{planckianBruin}
Bruin, J.A.N., Sakai, H., Perry, R.S., Mackenzie, A.P.: {Similarity of
  Scattering Rates in Metals Showing T-Linear Resistivity}. Science
  \textbf{339}(6121),  804--807 (2013). \doi{10.1126/science.1227612}

\bibitem{Eyring1}
Eyring, H.: The activated complex in chemical reactions. J. Chem. Phys.
  \textbf{3}, ~107 (1935). \doi{10.1063/1.1749604}

\bibitem{Eyring2}
Eyring, H.: Viscosity, plasticity, and diffusion as examples of absolute
  reaction rates. J. Chem. Phys.  \textbf{4}, ~283 (1936).
  \doi{10.1063/1.1749836}

\bibitem{Anup}
Gangopadhyay, A.K., Nussinov, Z., Kelton, K.F.: Quantum mechanical
  interpretation of the minimum viscosity of metallic liquids. Phys. Rev. E
  \textbf{106},  054150 (2022). \doi{10.1103/PhysRevE.106.054150}

\bibitem{Ramshaw}
Grissonnanche, G., Fang, Y., Legros, A., Verret, S., Laliberte, F., Collignon,
  C., Ataei, A., Dion, M., Zhou, J., Graf, D., Lawler, M.J., Goddard, P.,
  Taillefer, L., Ramshaw, B.J.: {Linear-in temperature resistivity from an
  isotropic Planckian scattering rate}. Nature  \textbf{595}, ~667 (2021).
  \doi{10.1038/s41586-021-03697-8}

\bibitem{KSS2}
Kovtun, P.K., Son, D.T., Starinets, A.O.: {Viscosity in Strongly Interacting
  Quantum Field Theories from Black Hole Physics}. Phys. Rev. Lett.
  \textbf{94},  111601 (2005). \doi{10.1103/PhysRevLett.94.111601}

\bibitem{Lieb_Robinson}
Lieb, E.H., Robinson, D.W.: {The finite group velocity of quantum spin
  systems}. Communications in Mathematical Physics  \textbf{28}(3),  251--257
  (1972). \doi{10.1007/BF01645779}

\bibitem{mssChaos}
Maldacena, J., Shenker, S.H., Stanford, D.: {A Bound on Chaos}. Journal of High
  Energy Physics  \textbf{2016}(8), ~106 (2016). \doi{10.1007/JHEP08(2016)106}

\bibitem{melting-speed}
Mousatov, C.H., Hartnoll, S.A.: {On the Planckian bound for heat diffusion in
  insulators}. Nature Physics  \textbf{16}(5),  579--584 (2020).
  \doi{10.1038/s41567-020-0828-6}

\bibitem{macroscopic}
Nussinov, Z.: Macroscopic length correlations in non-equilibrium systems and
  their possible realizations. Nuclear Physics B  \textbf{953},  114948 (2020).
  \doi{10.1016/j.nuclphysb.2020.114948}

\bibitem{nnbk-viscosity}
Nussinov, Z., Nogueira, F., Blodgett, M., Kelton, K.F.: {Thermalization and
  Possible Quantum Relaxation Times in ``Classical'' Fluids: Theory and
  Experiment}. arXiv:1409.1915v14. \url{https://arxiv.org/abs/1409.1915v14}

\bibitem{planckianAOP}
Nussinov, Z., Chakrabarty, S.: {Exact Universal Chaos, Speed Limit,
  Acceleration, Planckian Transport Coefficient, “Collapse” to Equilibrium,
  and Other Bounds in Thermal Quantum Systems}. Annals of Physics
  \textbf{443},  168970 (2022). \doi{10.1016/j.aop.2022.168970}

\bibitem{same_conjecture}
Schwarzhans, E., Binder, F.C., Huber, M., Lock, M.P.E.: Quantum measurements
  and equilibration: the emergence of objective reality via entropy
  maximisation. arXiv:2302.11253v1. \url{https://arxiv.org/abs/2302.11253v1}

\bibitem{taddeiQuantumSpeedLimit}
Taddei, M.M., Escher, B.M., Davidovich, L., de~Matos~Filho, R.L.: {Quantum
  Speed Limit for Physical Processes}. Phys. Rev. Lett.  \textbf{110},  050402
  (2013). \doi{10.1103/PhysRevLett.110.050402}

\bibitem{ktRev}
Trachenko, K.: Properties of condensed matter from fundamental physical
  constants. arXiv:2211.13342v1. \url{https://arxiv.org/abs/2211.13342v1}

\bibitem{KT}
Trachenko, K., Brazhkin, V.V.: {Minimal quantum viscosity from fundamental
  physical constants}. Science Advances  \textbf{6}(17) (2020).
  \doi{10.1126/sciadv.aba3747}

\bibitem{zoharViscosityPRR}
Xue, J., Nogueira, F.S., Kelton, K.F., Nussinov, Z.: Deviations from arrhenius
  dynamics in high temperature liquids, a possible collapse, and a viscosity
  bound. Phys. Rev. Res.  \textbf{4},  043047 (2022).
  \doi{10.1103/PhysRevResearch.4.043047}

\bibitem{zaanen2004}
Zaanen, J.: {Why the Temperature is High}. Nature  \textbf{430}(6999),
  512--513 (2004). \doi{10.1038/430512a}

\bibitem{jan-planck}
Zaanen, J.: {{Planckian dissipation, minimal viscosity and the transport in
  cuprate strange metals}}. SciPost Phys.  \textbf{6}, ~61 (2019).
  \doi{10.21468/SciPostPhys.6.5.061}

\end{thebibliography}
\end{document}